\documentclass[useAMS,usenatbib,letterpaper]{mn2e} 
\usepackage{times} 
\usepackage{amssymb}
\usepackage{lscape,graphicx}

\title[Galaxy morphologies and environment in the Abell 901/902
  supercluster from COMBO-17] {Galaxy morphologies and environment in
  the Abell 901/902 supercluster from COMBO-17}
 
\author [K. P. Lane et al]
{K. P. Lane$^1$\thanks{email: {\em ppxkl@nottingham.ac.uk}}, M. E. Gray$^1$,
A. Arag\'{o}n-Salamanca$^1$, C. Wolf$^2$, K. Meisenheimer $^3$\\ 1. School of Physics and Astronomy, The University of
Nottingham, University Park, Nottingham, NG7 2RD\\ 2. Department of Physics, Denys Wilkinson Bldg., University of Oxford, Keble Road,
Oxford OX1 3RH\\ 3. Max-Planck-Institut f\"{u}r Astronomie, K\"{o}nigstuhl 17, D-69117, Heidelberg, Germany}
 
\begin{document} 
 
\date{} 
\pagerange{\pageref{firstpage}--\pageref{lastpage}} \pubyear{} 
\maketitle 
\label{firstpage} 
 
\begin{abstract} 
We present a morphological study of galaxies in the A901/902 
supercluster from the COMBO-17 survey.  A total of 570 galaxies with
photometric redshifts in the range $0.155<z_{\rm
phot}<0.185$ are visually classified by three independent classifiers
to $M_V=-18$.  These morphological classifications
are compared to local galaxy density, distance from the nearest
cluster centre, local surface mass density from weak lensing, and
photometric classification.  At high local galaxy densities
$\log\Sigma_{10}$/Mpc$^2> 1.5$ a classical morphology-density
relation is found.  A correlation is also found between morphology and
local projected surface mass density, but no trend is observed with
distance to the nearest cluster.  This supports the finding that local
environment is more important to galaxy morphology than global cluster
properties.  The breakdown of the morphological catalogue by colour
shows a dominance of blue galaxies in the galaxies displaying
late-type morphologies and a corresponding dominance of red galaxies
in the early-type population.  Using the 17-band photometry from
COMBO-17, we further split the supercluster red sequence into old
passive galaxies and galaxies with young stars and dust according to
the prescription of \citet{Wolf_etal:2005}.  We find that the
dusty star-forming population describes an intermediate morphological
group between late-type and early-type galaxies, supporting the
hypothesis that field and group spiral galaxies are transformed into
S0s and, perhaps, ellipticals during cluster infall.
\end{abstract} 


\section{Introduction} 

\defcitealias{Gray_etal:2007}{Gray et al, in prep}

The precise role that environment plays in transforming the
morphological and star-formation properties of galaxies as they are
accreted onto groups and clusters remains unclear.  Well-known
correlations of cluster properties with environment hinting at
evolutionary effects include the cluster morphology-density relation
\citep{Dressler:1980, Dressler_etal:1997} and the increasing fraction
of blue galaxies in clusters at higher redshift
\citep{Butcher_Oemler_II:1978, Butcher_Oemler:1984}.  A galaxy's
encounters with other galaxies, with a hot intracluster medium (ICM)
or with a tidal cluster potential could all be pathways to
morphological alteration. However, it is also possible that galaxies in
the densest regions formed earlier, evolved more quickly, and thus
display more mature evolutionary states.

\citet{Dressler_etal:1997} find that at z $\sim$ 0.5 the fraction of
S0s is significantly lower than at present, with a proportional
increase in the spiral fraction. As the fraction of ellipticals is
already similar to that found locally, it is surmised that the low-z
S0 population formed chiefly from spirals.  One way to investigate the
effect of environment is to search for transitional objects in the
process of transformation.  For example, `passive spirals' displaying
spiral arms but no signatures of star formation could be identified as
an intermediary stage between spirals and S0s
\citep{Goto_etal:2003,Poggianti:2004}.  It is therefore important to
examine the variation of morphological populations with environment
(local mass, gas, and galaxy densities) to determine the physical
processes at work.

Historically both relaxed and irregular clusters have been the focus of morphological analysis. 
In fact \citet{Dressler:1980} showed that the morphology-density
relation is present in both. This implies that morphological
segregation is already in progress before cluster virialisation, if
heirarchical models of structure formation are to be believed. An
advanced stage of evolution will lead to cluster
properties being strongly correlated, and corresponding signatures of
different environmental influences may then be difficult to
untangle. In order to decouple the effects of mass density, galaxy
density, cluster-centric distance and gas content on galaxy
morphologies it is necessary to examine systems still in the process
of formation.  In a system that has not yet reached equilibrium, the
different components of the structure may still be segregated.

To this end we morphologically classify galaxies in
the Abell 901(a,b)/902 supercluster, a structure consisting of three
clusters and associated groups all within $5\times5$ Mpc at $z=0.17$.
We use imaging data from the COMBO-17 survey of this region
\citep{Wolf_etal:2001,Wolf_etal:2004}, which includes observations
from the ESO/MPG Wide-Field Imager in broad-band $UBVRI$ and a further
12 medium-band optical filters. The 17-band photometry provides
precision photometric redshifts (mean error $\sigma_z/(1+z) < 0.01$
for $R<20$ galaxies) and spectral energy
distributions (SEDs).  Additionally, a deep $R$-band image ($R<25.5$)
with 0.7\arcsec seeing provides excellent image quality for visual
classifications and weak gravitational lensing.  Further
multiwavelength coverage from X-ray to MIR of the
$0.5^\circ\times0.5^\circ$ field includes observations with
XMM-Newton, GALEX, Spitzer, a NIR extension to COMBO-17 using
Omega2000 on Calar Alto, plus spectra of the 300 brightest cluster
members from 2dF.  This extensive data set makes the supercluster a
uniquely well-positioned subject for detailed studies of galaxy
evolution.  The utility of this field will be greatly extended by a
forthcoming HST mosaic as part of the STAGES survey.

Additional interest comes from the fact that the supercluster is
dynamically complex, with no clear scaling relations between mass from
weak lensing maps \citep{Gray_etal:2002, Taylor_etal:2004}, galaxy
number density, velocity dispersion, and X-ray luminosity
\citepalias{Gray_etal:2007}.  This provides an excellent opportunity to
investigate which environmental properties have the greatest influence on
galaxy evolution, bearing in mind the important caveat that all such
quantities are seen in projection.  Here we investigate correlations
of visually classified galaxy morphologies with environment. Throughout we use a
concordance cosmology with $\Omega_m=0.27$,
$\Omega_\lambda=0.73$, and $H_0=71$ km s$^{-1}$ Mpc$^{-1}$ so that 1
arcmin = 168 kpc at the redshift of the supercluster.

\section{Visual galaxy classification} 
\label{Classif}

The cluster sample was chosen by a $0.155<z<0.185$ cut in photometric
redshift at an initial absolute magnitude cut of $M_V=-19$ (all magnitudes are
Vega).  According to
\citet{Wolf_etal:2005},  $\sigma_z < 0.01$ for this sample gives 99\%
completeness, if a Gaussian distribution is assumed. Using the same
redshift range but increasing the magnitude depth to $M_V=-18$
increases the error in the photo-z to $\sigma \approx 0.015$ and
therefore the completeness is $\sim$ 68\% at this luminosity
limit. The above sample of cluster galaxies is estimated to have
$\sim$ 60 non-cluster contaminants \citep{Wolf_etal:2005}. This represents 8.6\%
of our sample and so only introduces small uncertainties into our results.

Classifications were performed independently by three of the authors
(KPL, MEG, AAS) and combined in order to reduce classifier bias.
Galaxies were classified according to de Vaucouleurs' T-type scheme,
in which -5 is elliptical, -2 = S0, 1 = Sa, 2 = Sab, 3 = Sb, 4 = Sbc,
5 = Sc, 6 = Scd, 7 = Sd, 8 = Sdm, 9 = Sm/Irr
\citep{deVaucouleurs:1959}.  However, in terms of real ability to
discriminate by the classifiers this version of the T scale is
distorted: for example, a gap of three between E and S0s is far too
wide.  We therefore adopted an alternative T-type scale where the difference
between adjacent T-types is set to one and the gap between a given
Hubble type and its intermediary (e.g. Sa/Sab) is taken as a
difference of 0.5.  Additionally, the scheme allows combined
classifications in the case where a classifier cannot positively
separate two adjacent T-types (the most common example being E/S0 or
S0/E). The relative weighting assigned to the primary and secondary
choices is discussed below.  Comments were recorded if the morphology
was abnormal in any way and galaxies were flagged if they showed
visual signs of asymmetry, merger, or interaction.  

\begin{figure} 
\includegraphics[width=1\columnwidth]{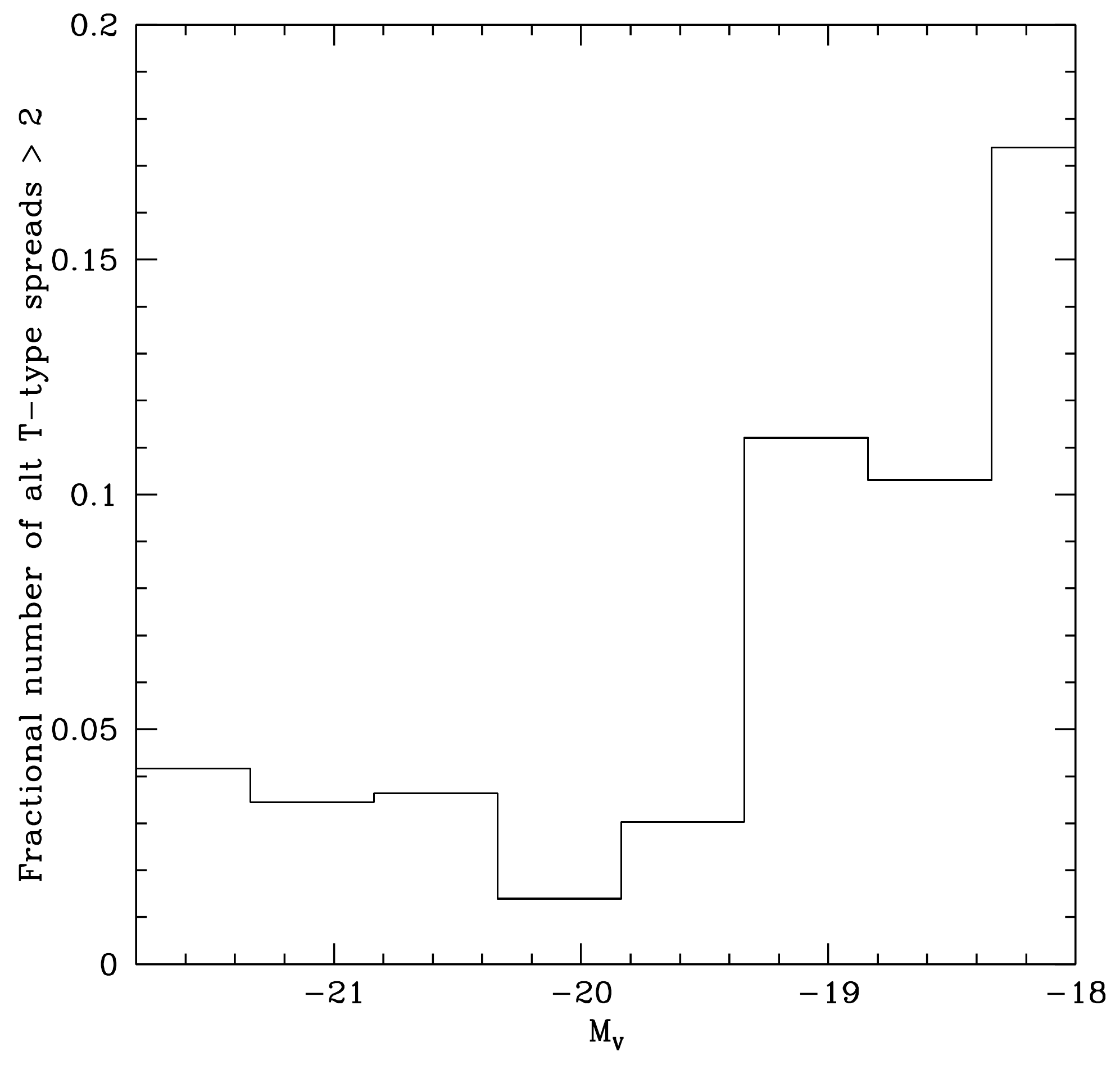}
\caption{Effects of magnitude on classification precision. The
 fraction of galaxies with a spread in classification of $>2$ alternative T-types from 3 independent classifiers is shown as a function of absolute
 magnitude. By $M_V=-18$ the fraction of galaxies with a spread in
 classifications $>2$ is almost 20\%.}
\label{fig-deltaT} 
\end{figure}

Combining the three classification sets gives some measure of the
reliability of the classifications.  Fig.~\ref{fig-deltaT} shows the
fraction of galaxies with a disagreement of more than two alternative T-types over
the three independent classifications.  The level of disagreement
increases with greater magnitude as the galaxies become fainter and
their structure harder to discern. Above an absolute magnitude of -18
the disagreement of $>2$ alternative T-types reaches $\sim 20\%$ of galaxies.  A
magnitude limit of $M_V=-18$ was therefore adopted for further
morphological analysis resulting in 570 galaxies.

These three classification sets were combined into one overall set
according to a set of rules based on the prescription used by the
EDisCS group \citep{Desai_etal:2006}. For each galaxy, the final
classification was computed by an equally weighted combination of the
three sets. In the case of a combined classification where a
classifier was unable to discriminate between adjacent T-types, the
primary and secondary classifications were assigned 3/4 and 1/4 of
that classifier's weighting, respectively. The T-type with the
highest combined weighting was taken as the final classification for
that galaxy. If two T-types had equal combined weightings the final
classification was randomly selected between them.  Finally, in the
case where there were more than two equally weighted types
the differences between the types were computed.  If one
difference was $\leq 3$ on the alternative T-type scale the final
classification was again chosen randomly between the two types
spanning this gap.  Otherwise if there was more than one gap of this
size or none $\leq 3$ then the median of the equally highly-weighted
T-types was taken as the final classification. The final Hubble type classification
catalog was comprised of 275 ellipticals, 126 S0, 59 Sa, 15 Sab, 49 Sb, 8 Sbc,
9 Sc, 2 Scd, 19 Sm/Irr and 6 unknown galaxies. For use in the following morphology-environment
studies spirals and Irr were binned together (167 in all), but given the small numbers it would make
no difference if we considered only pure spirals.

\begin{figure} 
\includegraphics[width=1\columnwidth]{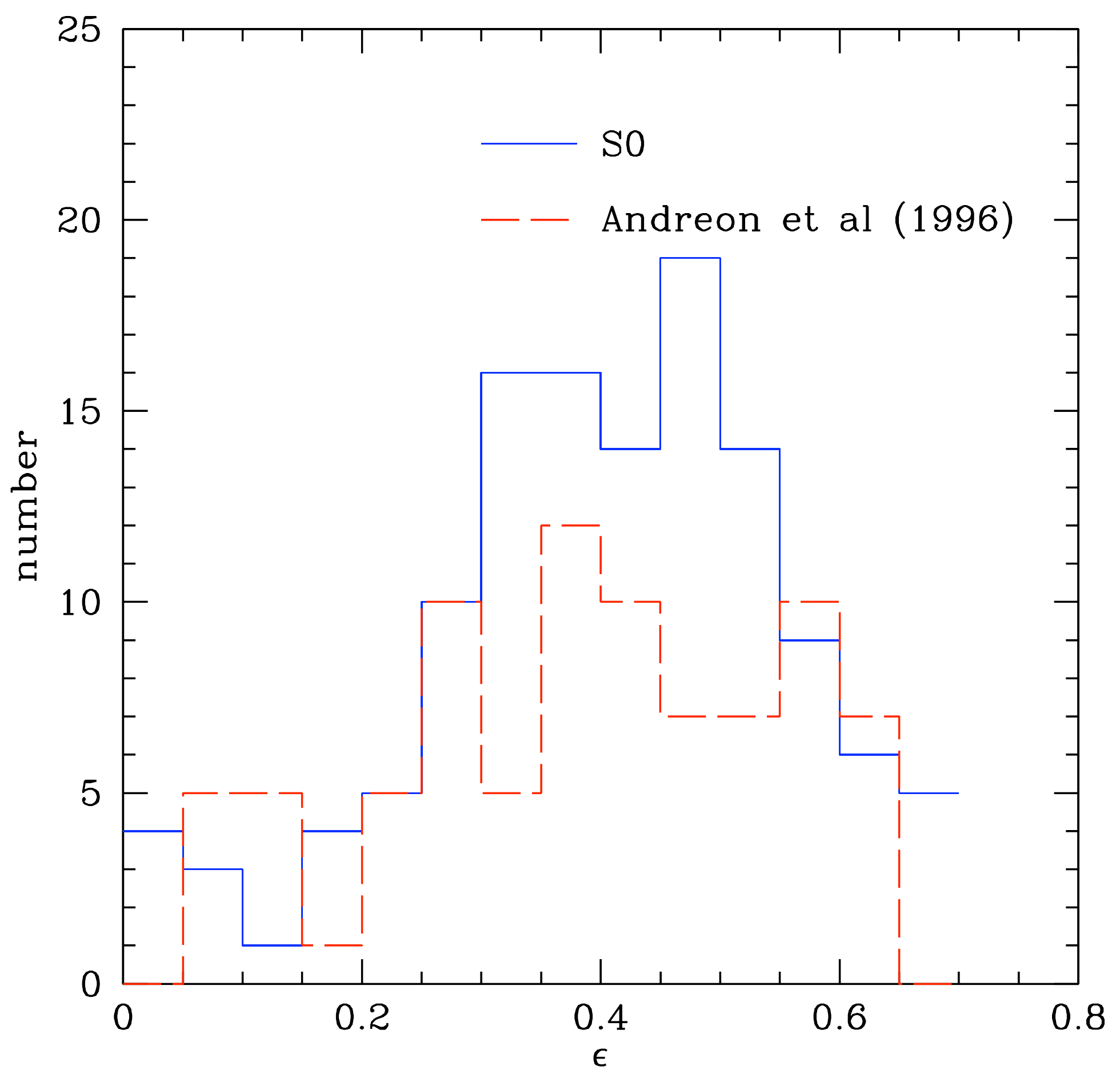}  
\includegraphics[width=1\columnwidth]{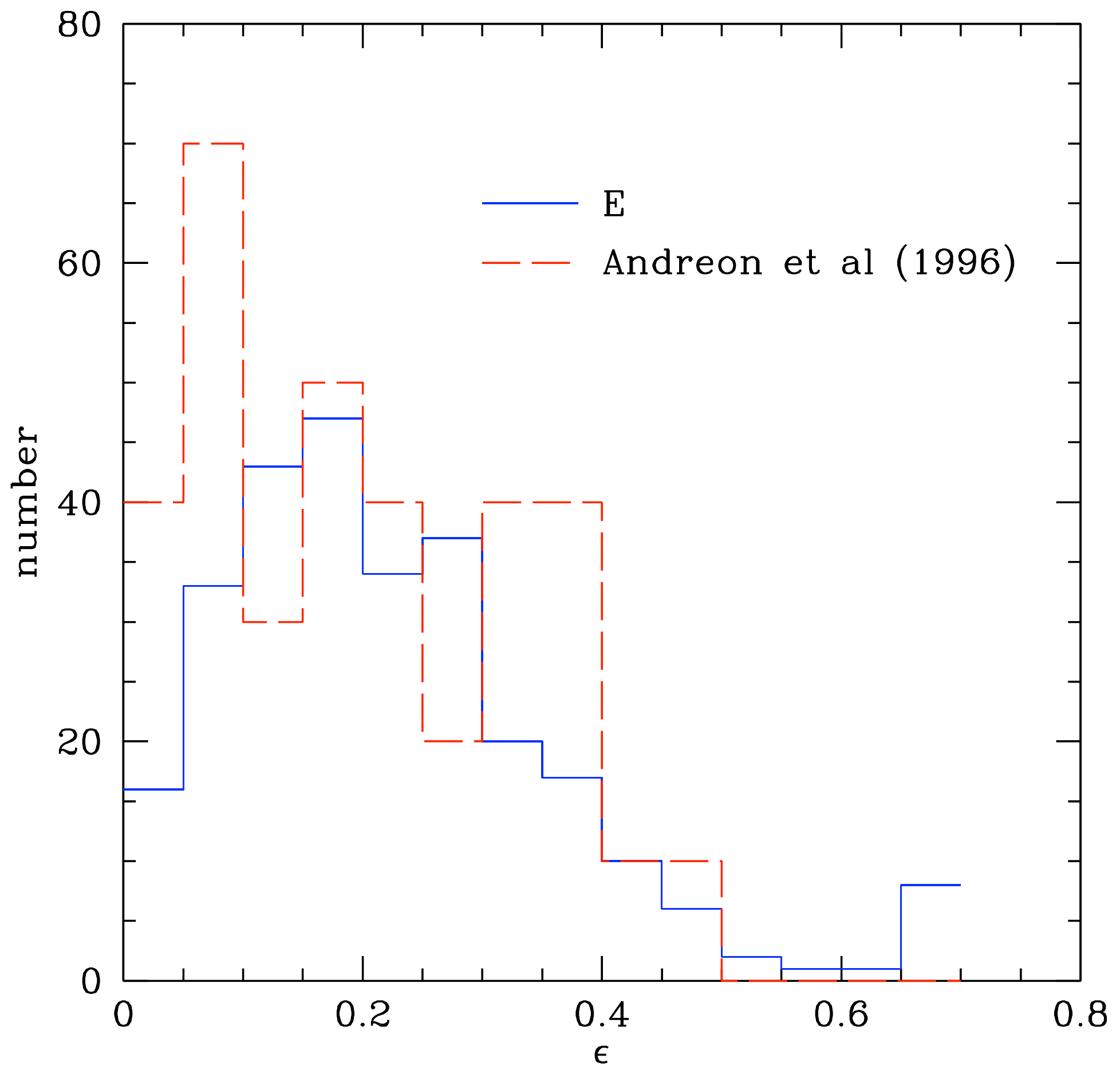}
\caption{Comparison of the ellipticities of the samples classified as
  Ellipticals and S0s in the A901/2 field with the ellipticities of
  the corresponding morphological samples as found in the Coma cluster
  \citep{Andreon_etal:1996}. The good agreement between the
  distributions of ellipticities in our samples and those in the Coma
  cluster, combined with the different ellipticity distributions of
  our Elliptical and S0 samples, provides some confidence in our
  ability to separate Elliptical and S0 classes.}

\label{fig-ellipticity} 
\end{figure}

The visual distinction between an S0 galaxy and an elliptical is
difficult, especially if the S0 is face on. Following
\citet{Dressler_etal:1997} we gauge the reliabilty of our
S0/E separation by comparing the ellipticities of our classified
galaxies to the ellipticities of galaxies in the
Coma cluster (\citealt{Andreon_etal:1996}, see Fig
~\ref{fig-ellipticity}). Ellipticities were measured from the $R$-band
image using SExtractor \citep{Bertin_Arnouts:1996}. Kolmogorov-Smirnov
tests give confidence intervals of 98.9\% for ellipticals and 53.1\%
for S0s being drawn from the same populations as their
\citet{Andreon_etal:1996} counterparts. This compares with an $\sim10^{-6}$ confidence that our ellipticals and S0s are drawn from the same
population. These results provide some confidence that we have reliably
separated these two classes.

\section{Linking morphology to cluster environment}
The morphology-density, or $\Sigma-T$, relation has been observed in a
large range of galaxy environments, from rich cluster cores, groups,
to field densities (e.g. \citealt{Postman_Geller:1984};
\citealt{Treu_etal:2003}).  Several physical mechanisms may be
responsible for transforming galaxy morphologies.  These will be
effective in different regimes: e.g. in groups where galaxy density
increases over the field, but where relative velocities are still low,
galaxy-galaxy interactions including major mergers may play a large
role \citep{Barnes:1992}.  The high velocities reached by galaxies in
the cores of rich clusters make mergers less likely, but increase the
efficacy of high-speed interactions such as harassment
\citep{Moore_etal:1999}.  Likewise, cluster-specific mechanisms such
as tidal stripping \citep{Merritt:1983} or the removal of a galaxy's
hot \citep{Larson_etal:1980} or cold \citep{Gunn_Gott:1972} gas by
ram-pressure processes require a steep potential or high ICM densities
and so are more likely to take effect in the inner regions of
clusters.

In light of the many physical mechanisms that may be at work in
different regimes, we examine galaxy morphology as a
function of several different proxies for `environment': local galaxy
density, projected mass density from lensing, and distance to the
nearest cluster centre.

\subsection{Linking morphology to local projected galaxy density}

The projected local galaxy density, $\Sigma_{10}$, was found by calculating the
area encompassed by the galaxy in question and its 9 nearest
neighbours. Only galaxies to $M_V=-18$ were used when calculating
$\Sigma_{10}$ as fainter galaxies will decrease completeness. 

Galaxies lying nearer to the edge of the image field than their 9th
nearest neighbour were removed from the catalogue, 37 in all.  This was to avoid
anomalously low densities resulting from missing neighbouring
galaxies outside the field-of-view.

Due to uncertainties in the photometric redshifts ($\sigma \approx
0.015$ at the magnitude limit used in this study, $M_V<-18$), we
estimate that $\sim 95$ potential field galaxies cannot be ruled out
as cluster members. This implies the minimum density we can measure is
$\sim 3.6$Mpc$^{-2}$. However, this is much smaller than the lowest
densities considered in our study ($\sim 25$Mpc$^{-2}$).

\begin{figure} 
\includegraphics[width=1\columnwidth]{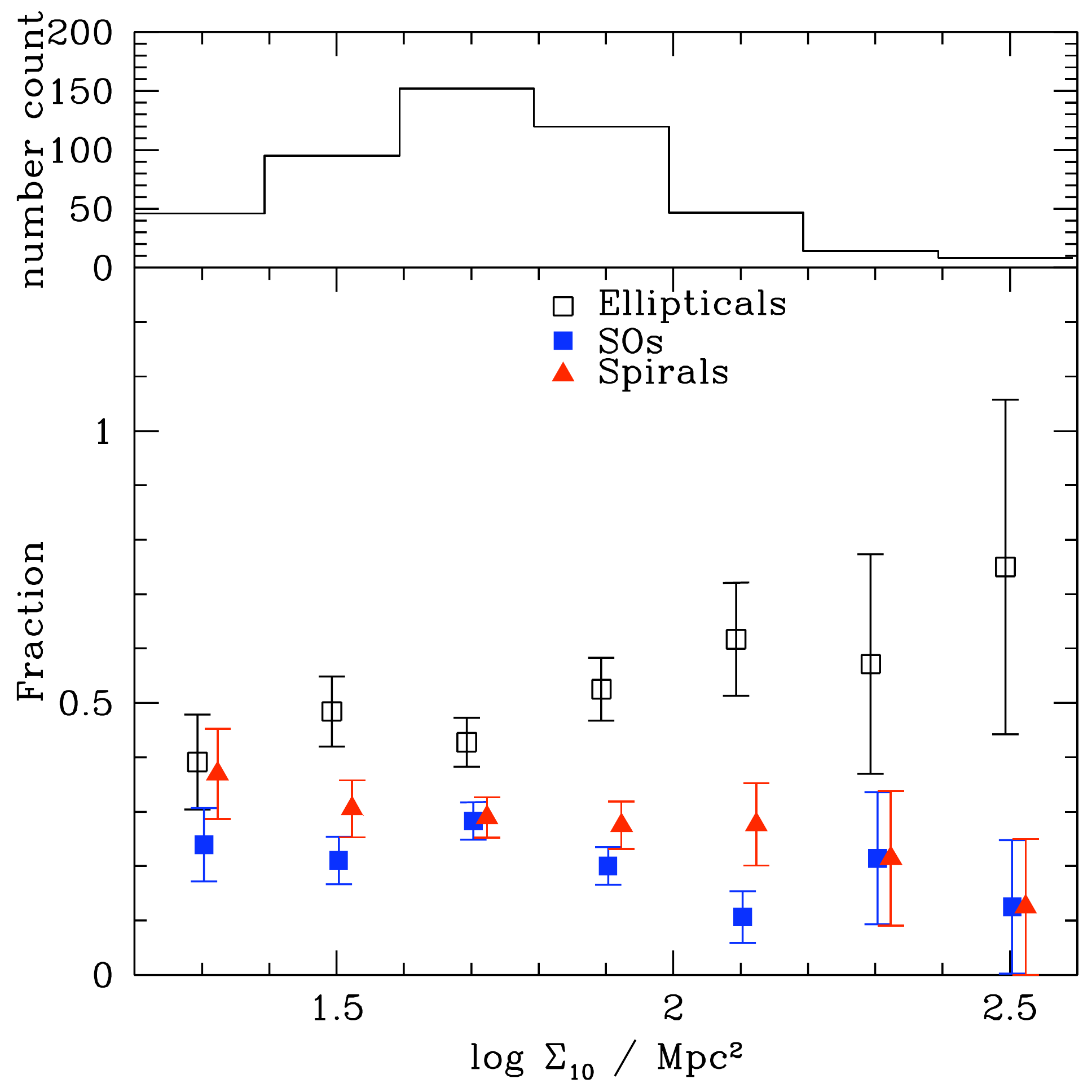}  
\caption{ Morphological type, as a fraction of the total, with
increasing local density. At high density there is a clear upward
trend in ellipticals and a downward trend in spirals. There appears to be no trend in S0s.  We ignored bins with fewer than 5 galaxies due to the
large sampling errors. This corresponds to log $\Sigma_{10} < 1.3$. The upper panel shows the total number of galaxies in each
density bin.}

\label{fig-MD} 
\end{figure}

The $\Sigma-T$ relation (Fig.~\ref{fig-MD}) shows a strong increase in
the fraction of ellipticals at high densities, a corresponding fall
in the fraction of spirals and the S0 fractions show no
correlation. This is the classical morphology-density
relation as seen in numerous other studies
\citep[e.g.][]{Oemler:1974,Dressler:1980,Dressler_etal:1997,Treu_etal:2003}.
Error bars are multinomial and were determined using Monte Carlo simulations.

By analysing data in the range $0<z<1$, including
\citet{Dressler:1980} at $z \sim 0$ and \citet{Dressler_etal:1997} at
$z \sim 0.5$, \citet{Smith_etal:2005} and \citet{Postman_etal:2005} find that the gradient of the early
type $\Sigma-T$ relation increases with lowering $z$ due to reducing numbers
of spirals and increasing numbers of S0s, especially in high density
regions. It is difficult to compare the results presented in Fig.~\ref{fig-MD}
with these studies because they use depths comparable to that used by
\citet{Dressler:1980}. To enable a comparison we cut our sample to the
same depth as used by \citet{Dressler:1980} ($M_V=-19.6$), after correcting for
differences in the cosmology used. Doing so reduces the number of
sources in our classified sample by more than a factor of 2 which increases the
error bars to point where no trends can be seen. However it is noted
that in our samples there are smaller fractions of S0s and larger fractions of Es when
compared with \citet{Dressler:1980}. This could be due to cluster to
cluster variation or systematic differences in classification.

If we compare our full sample with that of \citet{Dressler:1980}
relative to the cluster core $\Sigma_{10}$, the gradient in the $\Sigma-T$
relation for our elliptical population is shallower than found in the
highest density regions by \citet{Dressler:1980}. There are also
smaller fractions of S0s and larger fractions of spirals as
compared with \citet{Dressler:1980}. This is in agreement with
the findings of \citet{Smith_etal:2005} and
\citet{Postman_etal:2005}. At lower densities we find larger fractions of
ellipticals than \citet{Dressler:1980}. That our fractions of S0s are
so much smaller than \citet{Dressler:1980}, and that our elliptical
fractions are so much larger, is unexpected. However, S0 fraction does
vary by large amounts from cluster to cluster and the criteria used to
separate E and S0 classes varies from study to study. We also note that the
fraction of early types (elliptical + S0) in our
highest density bins is in keeping with the positive trend found in the fraction
of early type galaxies with decreasing redshift in high density
regions \citep{Smith_etal:2005}. Again it is noted that this
comparison is made using our full sample to $M_V=-18$ to enable sample
errors to be reduced to the point where trends can be seen.

With the adopted cosmology, 1 arcsecond corresponds to
$2.80$kpc at the cluster redshift. Similarly the seeing limited
resolution of our image is $1.96$kpc at the cluster redshift. This
compares with $\sim 700$pc for \citet{Postman_etal:2005}, using HST
ACS data, at $z \sim 1$, and $\sim 500$pc for
\citet{Dressler_etal:1997}, using HST WFPC2 data, at $z \sim
0.5$. Our resolution may result in later type spirals being
classified as earlier types since fine structure will be harder to
discern. This situation will be improved by using HST ACS data
obtained as part of the STAGES project.

\citet{Fasano_etal:2000} use ground based data with resolution 2-4 kpc
at the redshifts of their clusters ($0.1 \lesssim z \lesssim 0.25$). Again no comparisons can be drawn
when our sample is cut to the same absolute magnitude
($M_V=-20$). However, when comparing our full sample we find good
agreement between our MD relation and the MD relation found by
\citet{Fasano_etal:2000} in their high elliptical-concentration
clusters. Our data fits the trend of rising S0 fraction, falling
Sp fraction and no evolution in the fraction of E with redshift. We also
find our data to be consistent with the rising trends in the
$N_{S0}/N_{E}$ and $N_{S0}/N_{sp}$ fractions with lowering redshift, as
presented in \citet{Fasano_etal:2000}.

Note that the dynamic range in our $\Sigma_{10}$ measurements is
somewhat smaller than in the above studies, particularly at low
densities. However, this does not have a significant effect on the
comparisons discussed above.

\subsection{Linking morphology to projected cluster mass} 

As the mass of a cluster is made up predominantly of dark matter, one
might expect various environmental properties, such as local galaxy
density and the intracluster medium (ICM), to trace the potential
wells described by the dark matter mass of the cluster. This may well
be true in a virialised cluster where there is a positional degeneracy
between such environmental factors, however maps of projected mass and galaxy
distribution \citep{Gray_etal:2002} and extended X-ray emission \defcitealias{Gray_etal:2007}{Gray et al, in prep}
\citepalias{Gray_etal:2007} show that in the A901/902 supercluster such
scaling relations are not self-consistent from cluster to cluster. For example, Abell 901b
displays a prominent mass peak and $L_X = 2.35\times10^{44}$ erg
s$^{-1}$, yet is relatively deficient in galaxy numbers.  This then
provides an opportunity to ascertain if cluster mass has a direct
effect on galaxy morphology, or whether it is merely a tracer of other
morphology affecting environmental properties.

The projected surface mass density for this region was reconstructed
by \citet{Gray_etal:2002} from an analysis of weak lensing of faint
galaxies in the same COMBO-17 image.  The surface mass is measured as
a dimensionless quantity $\kappa$, where $\kappa = \Sigma/\Sigma_{\rm
crit}$ is the ratio of the projected surface mass density to the
critical surface mass density for lensing for a fixed source and lens
redshift.  For the supercluster lens at $z=0.17$ and a population of
faint lensed galaxies with $\langle z\rangle \sim 1$ we have
$\Sigma_{\rm crit} = 5.0\times10^{15}h M_{\sun}$ Mpc$^{-2}$.  The
\citet{Gray_etal:2002} map includes smoothing with a Gaussian of
$\sigma=60$ arcsec, and the rms noise in the map was estimated as $\sigma_\kappa=0.027$ through simulations.

\begin{figure} 
\includegraphics[width=1\columnwidth]{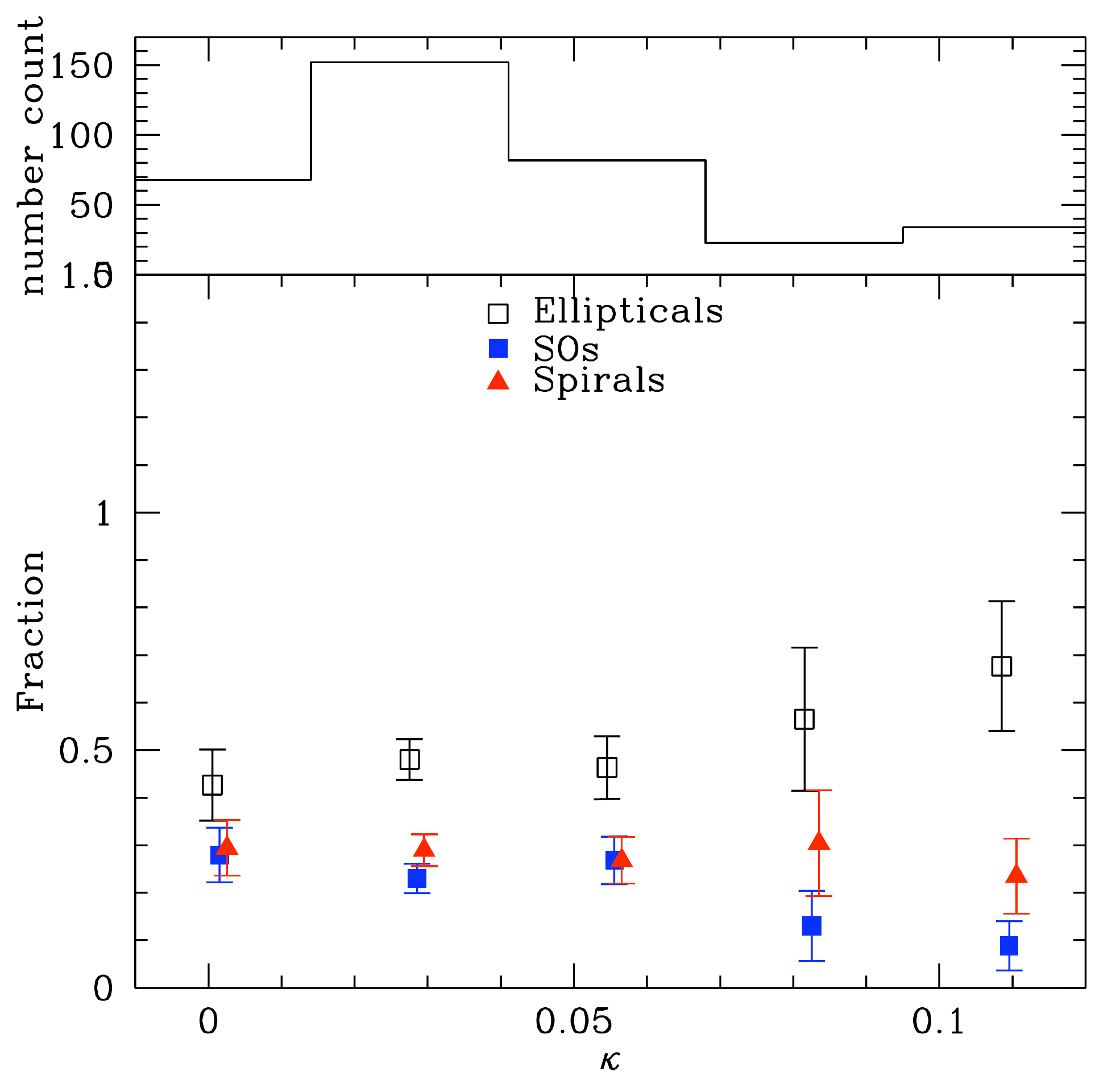}
\caption{ Morphological type, as a fraction of the total, with
increasing projected surface mass density, $\kappa$, from lensing. At
high surface density an upward trend in ellipticals and a downward
trend in S0s is observed. No clear trend is found in the spiral
population. The upper panel shows the total
number of galaxies in each $\kappa$ bin. The bin width corresponds to
the noise of the $\kappa$ maps, $\sigma_\kappa=0.027$.}
\label{fig-lens} 
\end{figure}

Fig.~\ref{fig-lens} shows a clear relation between projected mass and
morphology (a $\kappa-T$ relation) similar to the observed $\Sigma-T$
relation, but only at high mass densities (corresponding to the
$3\sigma$ $\kappa$ error regime).  An upward trend with $\kappa$ is found in
ellipticals and a downward trend in S0s, however spirals do not show any
correlation with projected mass. The different trends observed between
$\Sigma_{10}$ and $\kappa$ suggest that both are tracers of
morphology. Whether they are indepenent or, as is more likely, they
are tracers of different aspects of the same environmental driver for
morphology, remains to be determined. Fewer than $30\%$ of galaxies in
the highest $\kappa$ bins are classified as spirals.
This is analogous to the star-formation--$\kappa$ relation found for
the supercluster by \citet{Gray_etal:2004}, where the highest density
regions were found to be populated almost exclusively by galaxies with
quiescent SEDs.

\subsection{Linking morphology to cluster radius} 

Any correlation between clustercentric radius and the morphology of a
galaxy ($R-T$ relation) will most likely be a reflection of global
rather than local properties of the cluster since radius is not a
localised quantity. 

\begin{figure} 
\includegraphics[width=1\columnwidth]{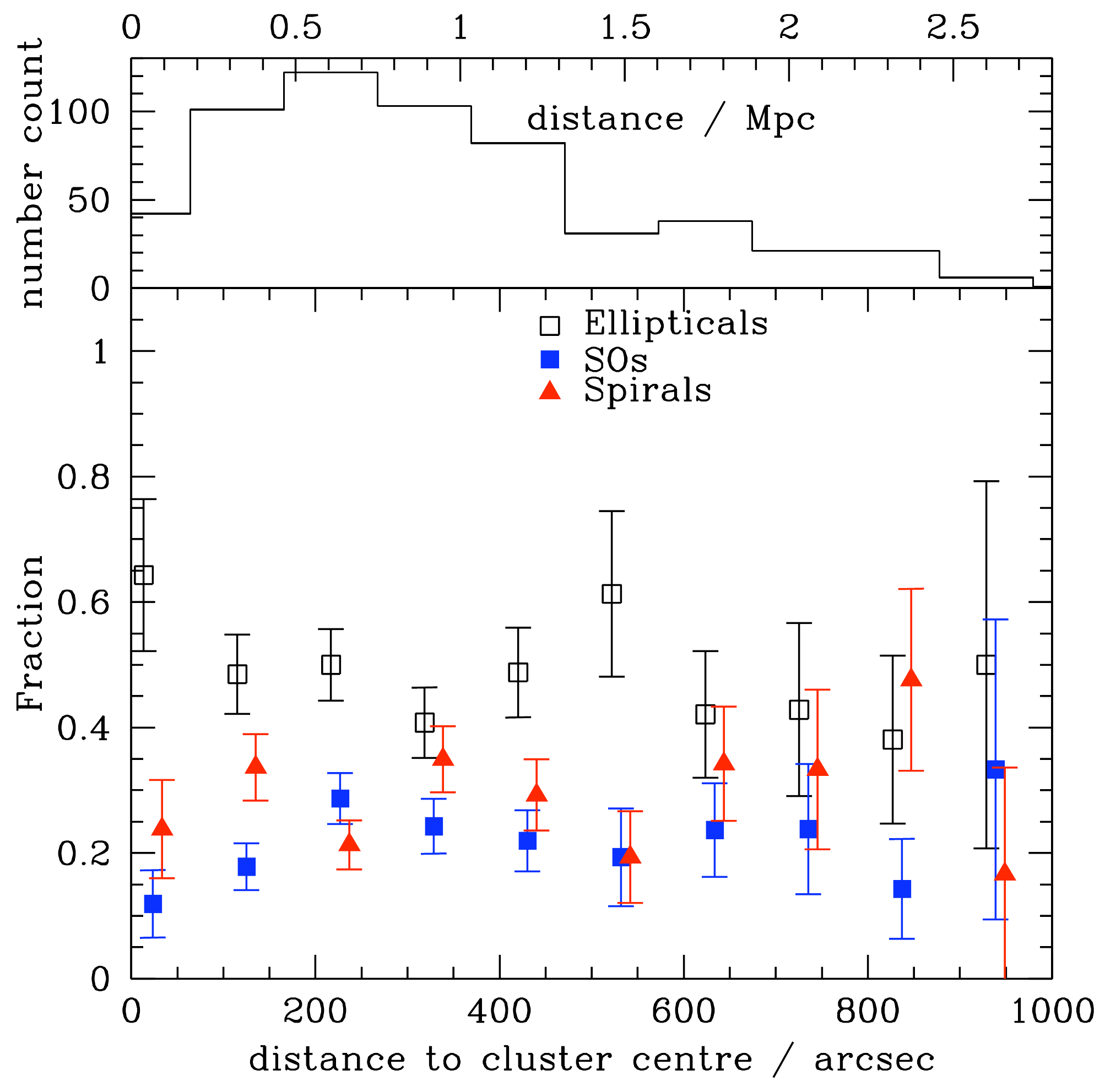}  
\caption{ Morphological type, as a fraction of the total, with
increasing distance to nearest cluster centre. No clear trend is
observed with increasing radial distance. Data points with zero sample
size have undetermined errors.}
\label{fig-radial} 
\end{figure}

The clustercentric radius is the distance to
the nearest cluster, where the cluster centre was defined as the peak
of the $\kappa$ map, although BCGs could have been used without much
change. Fig.~\ref{fig-radial} shows that there is no
clear trend between galaxy morphology and cluster radius, with only a
small rise in elliptical fractions and a small decrease in late-type
and S0 fractions at small radii. These small radii correspond to
regions which would be the least affected by the large scale cluster
merger. This result may then reflect the virialized cluster cores
where radius is degenerate with projected mass. 

The presence of a relation between morphology and local galaxy
density, combined with this apparent lack of a relation between
morphology and clustercentric radius adds further weight to the
hypothesis that local conditions have more effect on galaxy morphology
than global cluster properties.  Previous studies
\citep{Dressler_etal:1997} have shown that any radial dependence of
morphology is most likely a reflection of the $\Sigma-T$ relation due
to the one-to-one correspondence between density and radius in relaxed
clusters.  For the A901/902 system in particular the cluster
properties and radius are more decoupled than in relaxed systems due
to the dynamical complexity.  However, one caveat relevant to both
the $\Sigma-T$ and $R-T$ correlations is the possibility of
projection effects, particularly in the region between the A901a and
A901b clusters.  This possibility has been checked by masking out this region
and is found to have no appreciable effect on
the above results and hence is ruled out as a source of uncertainty.

\section{Linking morphology to photometric classification} 

In \citet{Wolf_etal:2005} the Abell 901/902 cluster galaxy sample
examined here was divided into three subpopulations: red, passively
evolving galaxies; blue star-forming galaxies; and a previously
unknown third population of red galaxies revealed by the 17-band
photometry.  This third population consists of cluster galaxies also
located along the cluster red sequence, but containing significant
amounts of young stars and dust (hereafter referred to as `dusty red
galaxies').

\begin{figure} 
\includegraphics[width=1\columnwidth]{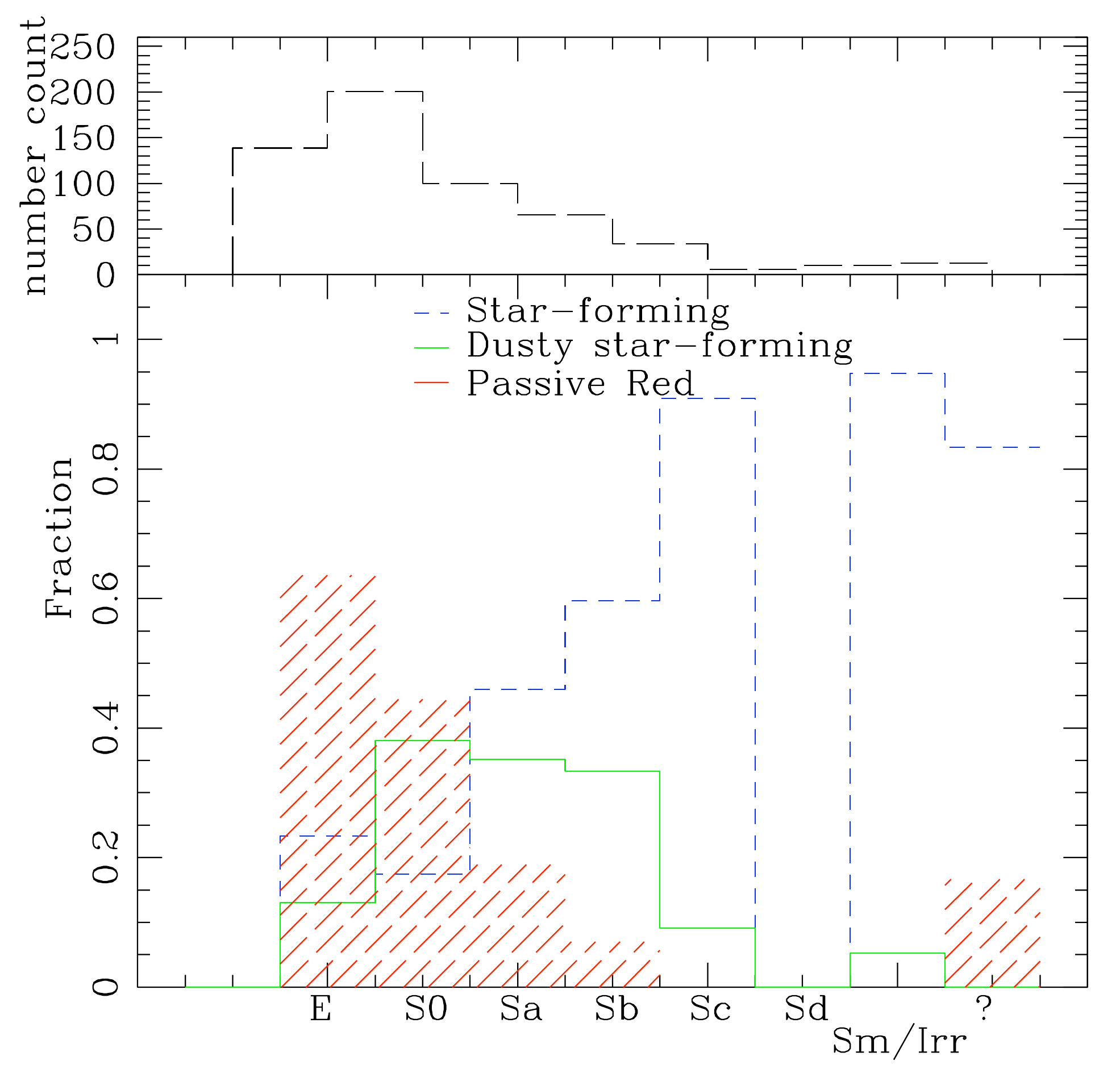}  
\caption{ The photometric colour of each morphological type as a
fraction of the total type population. Dusty red galaxies appear to
form an intermediate regime between star-forming late-type galaxies
and early-type passive galaxies. Data points with very large
sampling error have been omitted.}
\label{fig-colour} 
\end{figure}

Examining the photometric class of each morphological type,
Fig.~\ref{fig-colour} shows a clear trend.  The majority of
morphologically early-type galaxies (E,S0) are photometrically passive
and red.  Late-type galaxies (Sb,Sc) are predominantly blue.  The
third population of dusty red galaxies, on the other hand, shows a
distinct distribution of intermediate morphologies. In this case intermediate types
have been binned with the next highest integer alternative T-type for
clarity, e.g. Sab is binned with Sb, Sbc with Sc and so on.

Two possible origins for this intermediate population of dusty red
galaxies are posed by \citet{Wolf_etal:2005}. Firstly, that
they originate in the blue cloud and are in a state of being
transformed into red cluster galaxies or, secondly, that they are the
result of minor mergers of infalling galaxies with established cluster
galaxies.  To try and distinguish between these two formation scenarios the
merger or interaction state of each galaxy was noted. 

Of the classified galaxies which are photometrically found to be dusty
and red, only 14.2$\pm9.4\%$ were found by at least one classifier to
be in a state of interaction with a neighbouring galaxy or undergoing
a merger. This compares to 10.6$\pm6.5\%$ of the classified passive red
galaxies and 27.5$\pm8.5\%$ of the classified blue galaxies. Within
the uncertainties the three groups have consistent merger/interaction
fractions, however, the low fraction of interactions/mergers for dusty
red galaxies is inconsistent with a minor merger scenario for their
formation. However, it should be noted that a large fraction of mergers may go undetected
by a visual morphological analysis since this technique is only sensitive to
asymmetries or tidal features in morphology which may not be present in a
minor merger at scales larger than our resolution ($\sim
2$ kpc). Therefore the minor merger scenario for formation of dusty red
galaxies cannot be ruled out, but does look unlikely.

The dusty red galaxies represent a significant proportion of the
overall galaxy population (22.4\%) and do not appear to be a subset of
the blue or passive red galaxy populations.  The differences in
morphology are paralleled by differences in average spectra and
spatial distributions shown in \citet{Wolf_etal:2005}.  In
particular, they occupy regions of medium densities, avoiding high
densities nearer the cluster core as well as low density regions in
the cluster periphery \citep{Wolf_etal:2005}.  

These pieces of evidence would then suggest that one major route in
which infalling galaxies can be incorporated into the cluster is via
transformative processes that do not necessarily involve mergers.
Galaxies entering the cluster may have their star-formation ultimately
quenched, but after an initial phase of enhanced star-formation
\citep{Milvang-Jensen_etal:2003,Bamford_etal:2005}.  A
triggered starburst, possibly via interaction with the ICM, would
introduce dust via supernovae feedback to produce
the transitional dusty, red phase.  Ultimately the gas supply will be
exhausted and star formation quenched, leaving the remaining stars to
evolve passively on the red sequence.

\citet{Dressler_etal:1999} and \citet{Poggianti_etal:1999} find
similar spectroscopic populations of dusty starburst galaxies, or e(a) galaxies, at $z \sim 0.4 - 0.5$. They attribute these to the progenitors of
post-starburst k+a/a+k galaxies. For the COMBO-17 A901/2 field spectra
have been obtained for 64 cluster galaxies using the 2dF instrument (see
\citealt{Wolf_etal:2005}). The average spectra for dusty red galaxies
is seen to be inconsistent with that of k+a galaxies. The weak
[OII] emission as well as H$\delta$ absorption observed in the average
spectra of these dusty red galaxies are consistent with e(a) type
galaxies. To find such a large fraction (22.4\%) of potential k+a progenitor
galaxies at $z \sim 0.17$ is surprising given that
\citet{Dressler_etal:1999} find that 18\% of their
cluster sample exhibit k+a/a+k spectral types at $z \sim
0.5$. However, the continuing cluster-cluster merger seen in the
A901/2 system could well produce an increased incidence of e(a)
galaxies due to the large number of infalling galaxies.

\section{Conclusions} 

The complex dynamics present in the Abell 901/902 systems provide an
ideal testing ground for the distinction between the local and global
processes driving galaxy morphology.  In this paper we have examined
relations between visual galaxy morphologies and local measures of
environment including galaxy density, projected surface mass density
from lensing, and clustercentric radius.  The presence of a strong
$\Sigma-T$ and $\kappa-T$ relations and absence of a corresponding
$\Sigma-R$ relation shows that despite the large scale complexities of
the clusters, local conditions are still well correlated to
morphology.

Furthermore, the photometric breakdown of morphologies provides a
tantalising glimpse of galaxy transformation during infall, in action.
By determining the interaction and merger state of each classified
galaxy it was shown that the population of red sequence galaxies with
young stars and dust of \citet{Wolf_etal:2005} is not likely to be
explained by minor mergers of infalling galaxies with cluster members.
It is more plausible then that the dusty red galaxies are experiencing
additional, cluster-specific phenomena during infall, causing their
star-formation to become dust obscured and reddened. This would then
support a picture of cluster formation in which accreted galaxies can
be transformed through processes other than major or minor mergers,
most likely through induced star-bursts and associated dust
obscuration before the gas supply is stripped and/or exhausted, and
star-formation stops.

This study will be extended in the forthcoming STAGES survey of the A901/902
supercluster. The survey consists of an 80 orbit mosaic using HST/ACS
of the $0.5\times0.5$ degree region and will be combined with the
17-band photometric redshifts and other detailed multiwavelength
data sets. This mosaic will be used for morphological classifications not only for the bright end
of the cluster luminosity function probed here, but also the dwarf
galaxy population, which may be more sensitive to environmental processes
due to their lower gravitational potentials.  In this way we will
build up an even more detailed picture of galaxy evolution within a
complex environment.

\section*{Acknowledgments} 
KPL was supported by a PPARC studentship. MEG was supported by an Anne
McLaren Research Fellowship from the University of Nottingham. C. Wolf was supported by a PPARC Advanced fellowship. Thanks
go M. Merrifield and O. Almaini for useful and informative
discussions. We thank the anonymous referee for comments which greatly
improved the reliability of the results presented.

\bibliographystyle{mn2e}
\bibliography{mn-jour,papers_cited_by_KL}

\begin{thebibliography}{}

\bibitem[\protect\citeauthoryear{{Andreon}, {Davoust}, {Michard}, {Nieto} \&
  {Poulain}}{{Andreon} et~al.}{1996}]{Andreon_etal:1996}
{Andreon} S.,  {Davoust} E.,  {Michard} R.,  {Nieto} J.-L.,    {Poulain} P.,
  1996, \aaps, 116, 429

\bibitem[\protect\citeauthoryear{{Bamford}, {Milvang-Jensen},
  {Arag{\'o}n-Salamanca} \& {Simard}}{{Bamford}
  et~al.}{2005}]{Bamford_etal:2005}
{Bamford} S.~P.,  {Milvang-Jensen} B.,  {Arag{\'o}n-Salamanca} A.,    {Simard}
  L.,  2005, \mnras, 361, 109

\bibitem[\protect\citeauthoryear{{Barnes}}{{Barnes}}{1992}]{Barnes:1992}
{Barnes} J.~E.,  1992, \apj, 393, 484

\bibitem[\protect\citeauthoryear{{Bertin} \& {Arnouts}}{{Bertin} \&
  {Arnouts}}{1996}]{Bertin_Arnouts:1996}
{Bertin} E.,  {Arnouts} S.,  1996, \aaps, 117, 393

\bibitem[\protect\citeauthoryear{{Butcher} \& {Oemler}}{{Butcher} \&
  {Oemler}}{1978}]{Butcher_Oemler_II:1978}
{Butcher} H.,  {Oemler} A.,  1978, \apj, 226, 559

\bibitem[\protect\citeauthoryear{{Butcher} \& {Oemler}}{{Butcher} \&
  {Oemler}}{1984}]{Butcher_Oemler:1984}
{Butcher} H.,  {Oemler} A.,  1984, \apj, 285, 426

\bibitem[\protect\citeauthoryear{{de Vaucouleurs}}{{de
  Vaucouleurs}}{1959}]{deVaucouleurs:1959}
{de Vaucouleurs} G.,  1959, Handbuch der Physik, 53, 275

\bibitem[\protect\citeauthoryear{{Desai}, {Dalcanton}, {Arag\'on-Salamanca},
  {Jablonka}, {Poggianti}, {Gogarten}, {Simard}, {Clowe}, {Halliday},
  {Milvang-Jensen} \& {Pell\'o}}{{Desai} et~al.}{2006}]{Desai_etal:2006}
{Desai} V.,  {Dalcanton} J.~J.,  {Arag\'on-Salamanca} A.,  {Jablonka} P.,
  {Poggianti} B.,  {Gogarten} S.~M.,  {Simard} L.,  {Clowe} D.,  {Halliday} C.,
   {Milvang-Jensen} B.,    {Pell\'o} R.,  2006, \apj, p. submitted 2006

\bibitem[\protect\citeauthoryear{{Dressler}}{{Dressler}}{1980}]{Dressler:1980}
{Dressler} A.,  1980, \apj, 236, 351

\bibitem[\protect\citeauthoryear{{Dressler}, {Oemler}, {Couch}, {Smail},
  {Ellis}, {Barger}, {Butcher}, {Poggianti} \& {Sharples}}{{Dressler}
  et~al.}{1997}]{Dressler_etal:1997}
{Dressler} A.,  {Oemler} A.~J.,  {Couch} W.~J.,  {Smail} I.,  {Ellis} R.~S.,
  {Barger} A.,  {Butcher} H.,  {Poggianti} B.~M.,    {Sharples} R.~M.,  1997,
  \apj, 490, 577

\bibitem[\protect\citeauthoryear{{Dressler}, {Smail}, {Poggianti}, {Butcher},
  {Couch}, {Ellis} \& {Oemler}}{{Dressler} et~al.}{1999}]{Dressler_etal:1999}
{Dressler} A.,  {Smail} I.,  {Poggianti} B.~M.,  {Butcher} H.,  {Couch} W.~J.,
  {Ellis} R.~S.,    {Oemler} A.~J.,  1999, \apjs, 122, 51

\bibitem[\protect\citeauthoryear{{Fasano}, {Poggianti}, {Couch}, {Bettoni},
  {Kj{\ae}rgaard} \& {Moles}}{{Fasano} et~al.}{2000}]{Fasano_etal:2000}
{Fasano} G.,  {Poggianti} B.~M.,  {Couch} W.~J.,  {Bettoni} D.,
  {Kj{\ae}rgaard} P.,    {Moles} M.,  2000, \apj, 542, 673

\bibitem[\protect\citeauthoryear{{Goto}, {Okamura}, {Sekiguchi}, {Bernardi},
  {Brinkmann}, {G{\'o}mez}, {Harvanek}, {Kleinman}, {Krzesinski}, {Long},
  {Loveday}, {Miller}, {Neilsen}, {Newman}, {Nitta}, {Sheth}, {Snedden} \&
  {Yamauchi}}{{Goto} et~al.}{2003}]{Goto_etal:2003}
{Goto} T.,  {Okamura} S.,  {Sekiguchi} M.,  {Bernardi} M.,  {Brinkmann} J.,
  {G{\'o}mez} P.~L.,  {Harvanek} M.,  {Kleinman} S.~J.,  {Krzesinski} J.,
  {Long} D.,  {Loveday} J.,  {Miller} C.~J.,  {Neilsen} E.~H.,  {Newman} P.~R.,
   {Nitta} A.,  {Sheth} R.~K.,  {Snedden} S.~A.,    {Yamauchi} C.,  2003,
  \pasj, 55, 757

\bibitem[\protect\citeauthoryear{{Gray}}{{Gray}}{2007}]{Gray_etal:2007}
{Gray} M.~E.,  2007, in prep

\bibitem[\protect\citeauthoryear{{Gray}, {Taylor}, {Meisenheimer}, {Dye},
  {Wolf} \& {Thommes}}{{Gray} et~al.}{2002}]{Gray_etal:2002}
{Gray} M.~E.,  {Taylor} A.~N.,  {Meisenheimer} K.,  {Dye} S.,  {Wolf} C.,
  {Thommes} E.,  2002, \apj, 568, 141

\bibitem[\protect\citeauthoryear{{Gray}, {Wolf}, {Meisenheimer}, {Taylor},
  {Dye}, {Borch} \& {Kleinheinrich}}{{Gray} et~al.}{2004}]{Gray_etal:2004}
{Gray} M.~E.,  {Wolf} C.,  {Meisenheimer} K.,  {Taylor} A.,  {Dye} S.,  {Borch}
  A.,    {Kleinheinrich} M.,  2004, \mnras, 347, L73

\bibitem[\protect\citeauthoryear{{Gunn} \& {Gott}}{{Gunn} \&
  {Gott}}{1972}]{Gunn_Gott:1972}
{Gunn} J.~E.,  {Gott} J.~R.~I.,  1972, \apj, 176, 1

\bibitem[\protect\citeauthoryear{{Larson}, {Tinsley} \& {Caldwell}}{{Larson}
  et~al.}{1980}]{Larson_etal:1980}
{Larson} R.~B.,  {Tinsley} B.~M.,    {Caldwell} C.~N.,  1980, \apj, 237, 692

\bibitem[\protect\citeauthoryear{{Merritt}}{{Merritt}}{1983}]{Merritt:1983}
{Merritt} D.,  1983, \apj, 264, 24

\bibitem[\protect\citeauthoryear{{Milvang-Jensen}, {Arag{\' o}n-Salamanca},
  {Hau}, {J{\o}rgensen} \& {Hjorth}}{{Milvang-Jensen}
  et~al.}{2003}]{Milvang-Jensen_etal:2003}
{Milvang-Jensen} B.,  {Arag{\' o}n-Salamanca} A.,  {Hau} G.~K.~T.,
  {J{\o}rgensen} I.,    {Hjorth} J.,  2003, \mnras, 339, L1

\bibitem[\protect\citeauthoryear{{Moore}, {Lake}, {Quinn} \& {Stadel}}{{Moore}
  et~al.}{1999}]{Moore_etal:1999}
{Moore} B.,  {Lake} G.,  {Quinn} T.,    {Stadel} J.,  1999, \mnras, 304, 465

\bibitem[\protect\citeauthoryear{{Oemler}}{{Oemler}}{1974}]{Oemler:1974}
{Oemler} A.~J.,  1974, \apj, 194, 1

\bibitem[\protect\citeauthoryear{{Poggianti}}{{Poggianti}}{2004}]{Poggianti:20%
04}
{Poggianti} B.,  2004, in Baryons in Dark Matter Halos {Evolution of galaxies
  in clusters}

\bibitem[\protect\citeauthoryear{{Poggianti}, {Smail}, {Dressler}, {Couch},
  {Barger}, {Butcher}, {Ellis} \& {Oemler}}{{Poggianti}
  et~al.}{1999}]{Poggianti_etal:1999}
{Poggianti} B.~M.,  {Smail} I.,  {Dressler} A.,  {Couch} W.~J.,  {Barger}
  A.~J.,  {Butcher} H.,  {Ellis} R.~S.,    {Oemler} A.~J.,  1999, \apj, 518,
  576

\bibitem[\protect\citeauthoryear{{Postman}, {Franx}, {Cross}, {Holden}, {Ford},
  {Illingworth}, {Goto} \& {others}}{{Postman}
  et~al.}{2005}]{Postman_etal:2005}
{Postman} M.,  {Franx} M.,  {Cross} N.~J.~G.,  {Holden} B.,  {Ford} H.~C.,
  {Illingworth} G.~D.,  {Goto} T.,    {others} 2005, \apj, 623, 721

\bibitem[\protect\citeauthoryear{{Postman} \& {Geller}}{{Postman} \&
  {Geller}}{1984}]{Postman_Geller:1984}
{Postman} M.,  {Geller} M.~J.,  1984, \apj, 281, 95

\bibitem[\protect\citeauthoryear{{Smith}, {Treu}, {Ellis}, {Moran} \&
  {Dressler}}{{Smith} et~al.}{2005}]{Smith_etal:2005}
{Smith} G.~P.,  {Treu} T.,  {Ellis} R.~S.,  {Moran} S.~M.,    {Dressler} A.,
  2005, \apj, 620, 78

\bibitem[\protect\citeauthoryear{{Taylor}, {Bacon}, {Gray}, {Wolf},
  {Meisenheimer}, {Dye}, {Borch}, {Kleinheinrich}, {Kovacs} \&
  {Wisotzki}}{{Taylor} et~al.}{2004}]{Taylor_etal:2004}
{Taylor} A.~N.,  {Bacon} D.~J.,  {Gray} M.~E.,  {Wolf} C.,  {Meisenheimer} K.,
  {Dye} S.,  {Borch} A.,  {Kleinheinrich} M.,  {Kovacs} Z.,    {Wisotzki} L.,
  2004, \mnras, 353, 1176

\bibitem[\protect\citeauthoryear{{Treu}, {Ellis}, {Kneib}, {Dressler}, {Smail},
  {Czoske}, {Oemler} \& {Natarajan}}{{Treu} et~al.}{2003}]{Treu_etal:2003}
{Treu} T.,  {Ellis} R.~S.,  {Kneib} J.,  {Dressler} A.,  {Smail} I.,
  {Czoske} O.,  {Oemler} A.,    {Natarajan} P.,  2003, \apj, in press
  (astro-ph/0303267)

\bibitem[\protect\citeauthoryear{{Wolf}, {Dye}, {Kleinheinrich},
  {Meisenheimer}, {Rix} \& {Wisotzki}}{{Wolf} et~al.}{2001}]{Wolf_etal:2001}
{Wolf} C.,  {Dye} S.,  {Kleinheinrich} M.,  {Meisenheimer} K.,  {Rix} H.-W.,
  {Wisotzki} L.,  2001, \aap, 377, 442

\bibitem[\protect\citeauthoryear{{Wolf}, {Gray} \& {Meisenheimer}}{{Wolf}
  et~al.}{2005}]{Wolf_etal:2005}
{Wolf} C.,  {Gray} M.~E.,    {Meisenheimer} K.,  2005, \aap, 443, 435

\bibitem[\protect\citeauthoryear{{Wolf}, {Meisenheimer}, {Kleinheinrich},
  {Borch}, {Dye}, {Gray}, {Wisotzki}, {Bell}, {Rix}, {Cimatti}, {Hasinger} \&
  {Szokoly}}{{Wolf} et~al.}{2004}]{Wolf_etal:2004}
{Wolf} C.,  {Meisenheimer} K.,  {Kleinheinrich} M.,  {Borch} A.,  {Dye} S.,
  {Gray} M.,  {Wisotzki} L.,  {Bell} E.~F.,  {Rix} H.-W.,  {Cimatti} A.,
  {Hasinger} G.,    {Szokoly} G.,  2004, \aap, 421, 913

\end{thebibliography}

\label{lastpage} 

\end{document}